\documentstyle[12pt]{article}

\begin{document}
\vskip2cm

\begin{center}{\large \bf Positivity Constraints on Chiral
Perturbation Theory Pion-Pion Scattering Amplitudes}
\end{center}
\vskip .21cm
\begin{center}{Petre Di\c t\u a\footnote{email: dita@theor1.theory.ifa.ro}}
\end{center}
\vskip .21cm
\begin{center}
{National Institute of Physics \& Nuclear Engineering}\\
{Bucharest, PO Box MG6, Romania}
\end{center}
\vskip.71cm
\begin{abstract}
We test the positivity property of the chiral perturbation theory (ChPT)
pion-pion scattering amplitudes within the Mandelstam triangle. In the 
one-loop approximation, ${\cal O}(p^4)$, the positivity constrains only the coefficients
$b_3$ and $b_4$, namely one obtains that $b_4$ and the linear combination
$b_3+3 b_4 $ are positive quantities. The two-loops approximation gives inequalities
involving all the six arbitrary parameters entering ChPT amplitude, but the corrections
to the one-loop approximation  results are small. ChPT amplitudes pass unexpectedly well all
the positivity tests giving strong support to
 the idea that ChPT is the good theory of the low energy pion-pion scattering.
\vskip3mm
\centerline{PACS: 13.75.Lb, 12.39.Fe, 11.55.Fv, 11.30.Rd}
Key-words: Pion-Pion Scattering, Chiral Perturbation Theory, Positivity Constraints
\end{abstract}
\vskip1.cm
\section{Introduction}
Chiral perturbation theory  (ChPT) is considered  a low-energy
effective approximation of QCD, in particular it provides a
representation of the elastic pion-pion scattering amplitudes
that is crossing symmetric and has good analyticity properties.
In a seminal paper\cite{GL} Gasser and Leutwyler developed
ChPT
which allows one to compute many Green functions
involving
low-energy pions.  It is well known that the physical
pion-pion scattering amplitudes can be expressed in terms of a
single function $A(s,t,u)$ whose form was obtained as a series
expansion in powers  of the external momenta and of the
light
quark masses.  The first term of the series was given by
Weinberg
\cite{We}, the second by Gasser and Leutwyler
\cite{GL} and only
recently  a two-loop calculation\cite{Kn,Co}
was obtained. In this approximation the function $A(s,t,u)$ has 
the following form 
$$A(s,t,u)=
a(s-1)+a^2\left[b_1
+b_2s+b_3s^2+b_4(t-u)^2\right]+$$
$$a^2\left[F^{(1)}(s)
+G^{(1)}(s,t)+G^{(1)}(s,u)\right]+
a^3\left[b_5s^3+b_6s(t-u)^2\right]
+\eqno(1.1)$$
$$a^3\left[F^{(2)}(s)+G^{(2)}(s,t)+G^{(2)}(s,u)\right]+{\cal O}(a^4)$$
where $a=(M_{\pi}/F_{\pi})^2$, $M_{\pi}$ is the mass of the
physical
pion, $F_{\pi}$ the pion decay constant, $s, t,u$ are
the usual
Mandelstam variables, expressed in units of the physical 
pion
mass squared $M_{\pi}^2$
$$s=(p_1+p_2)^{2}/M_{\pi}^{2},\,\,\,
t=(p_1-p_2)^{2}/M_{\pi}^{2},\,\,\, u=(p_1-p_3)^{2}/M_{\pi}^{2},$$
$F^{(i)}(s)$ and $G^{(i)}(s,t)$ are known functions and
$b_i,\,\,i=1,\dots,6$ are arbitrary parameters which cannot
be determined by ChPT \cite{GL,Kn,Co}.
In any realistic
comparison with experiment we have  to provide some
numerical values
for all these parameters obtained from other sources.
 One hopes that by
using unitarity
this can be done although, until now, no
program for
implementing this property has been presented.  
The common
belief
is that imposing unitarity is not a simple matter since
its
implementation in one channel destroys crossing symmetry
in
other channels. However there is a weak form of unitarity,
the
positivity of the absorptive parts, which is a linear
property
and which can be imposed. This property was used
thirty
years ago to obtain constraints on the $\pi^0\pi^0$\,
$s$-wave
partial amplitude, $f_0(s)$, in the unphysical region
$0<s<4$
and on the $d$-wave scattering lengths. These
constraints were
useful because at that time almost nothing was
known about the
explicit form of the scattering amplitudes and
they were used
in testing models for pion-pion partial-wave
amplitudes.  The
advantage of ChPT is that it furnishes an
explicit form for the
pion-pion scattering amplitudes whose
unknown part  is
contained in a few numerical coefficients.
Thus is of certain
interest to see how these properties reflect
into constraints on
the $b_i$ coefficients entering
Eq.(1.1).

  Beginning with the paper
\cite{M1}, Martin has
used the positivity, analyticity and crossing 
symmetry to
obtain
constraints on the $\pi^0\pi^0$  s-wave partial amplitude, $f_0(s)$,
in the
unphysical region $0<s<4$; a few of them have the following form
\cite{M2}
$$f_0(4)>f_0(0)>f_0(3.15)\qquad
f_0(0)>{1\over
2}\int_2^4f_0(s)\, ds$$ $$F(0,2(1+{1\over{
\sqrt
3}}))>f_0(2(1+{1\over \sqrt 3}))\eqno(1.2)$$
\noindent where $F(s,t)$ denotes the $\pi^0\pi^0$ elastic
scattering
amplitude.  A more complete set is found in ref.
\cite{Mr}.
The most elaborate form of these constraints is the
following
result: the $\pi^0\pi^0$ $s$-wave,
$f_0(s)$,
has a minimum located between
$1.218989<s<1.696587$\,
\cite{M2,Mr,AC,GA,CP,Gr,
Di} and this
result can be improved only by unitarity. 
 These results can
be translated into constraints on
the parameters $b_i$ entering
ChPT pion-pion scattering
amplitudes.  As we will see later all
the above inequalities are equivalent
in the one-loop
approximation, ${\cal O}(p^4)$,  to a single constraint on the
coefficients $b_i$ whose typical form
is
$$b_3+3 b_4\ge {37\over 1920 \pi^2}$$
the only difference
being the numerical value appearing on the right hand side. This is
 the explanation of the inefficacy of these apparently
distinct constraints which
was observed from the beginning by
people constructing models for pion-pion
partial-waves.

In
this paper we work only in the unphysical
region $|s,t,u|<4$, i.e. 
there where the amplitude (1.1) is considered to be
a
very good approximation to the true amplitude.  
 Other
approaches use information from the physical region to obtain
constraints on the same parameters $b_i$\, \cite{Kn1,Bi,Wa}.
In the one-loop approximation, ${\cal O}(p^4)$,
the positivity
property constrains only the coefficients 
$b_3$ and $b_4$.
By taking into 
 account the ${\cal O}(p^6)$ contributions one gets
constraints
involving 
 all
 the six 
parameters
entering
Eq.(1.1).  

By using the unitarity bounds  on
$\pi^0\pi^0$ scattering amplitude in the
unphysical region
\cite{LM} one get upper and lower bounds on some
linear
combinations of the parameters $b_i$. These unitarity bounds are not 
very constraining; to see this we give that one
 obtained from the bound on $F(2,0)$,
where as above $F(s,t)$ denotes the $\pi^0\pi^0$ amplitude.
The bound is  $-3.5\le F(2,0)\le 2.9$ and it is equivalent to the following
lower and upper bounds
$$-3.5\times 32\pi\le a+ a^2\left(3 b_1+4b_2+8b_3+8b_4+{9\over 32\pi^2}\right)
+a^3\left[16b_5+16b_6+\right.$$
$$\left.{(4-\pi)\over\pi^2}
\left({5b_1\over 16}+
{b_2\over 2}+{11 b_3\over 12}+{5b_4\over 12}\right)+
{965\over 3456\pi^4}-{251\over 3456\pi^3}+{41\over 6144\pi^2}\right]\le 2.9
\times 32\pi$$
Because of the factor $32 \pi$ appearing on left and right hand side
 the bounds are not very strong and we will not consider them here.

 The physical
isospin amplitudes $F^I$
can be expressed in terms of the
single function $A(s,t,u)$ as follows

$$
\begin{array}{lcl}F^0(s,t,u)& =& 3
A(s,t,u)+A(t,u,s)+A(u,s,t)\\
F^1(s,t,u)&=&A(t,u,s)-A(u,s,t)\\
F^2(s,t,u)&=&A(t,u,s)+A(u,s,t)
\end{array}$$
where $A(s,t,u)$
is given by Eq.(1.1).

 Having only three
independent
amplitudes one gets only three independent
constraints since the crossing
symmetry is an exact symmetry
for the ChPT amplitudes. The 
construction 
of
our positivity
constraints 
is outlined 
in 
the next Section 
where
we
present an over-determined 
system of constraints.
 Their
implications on the
coefficients 
$b_i$ 
are discussed 
in
Section III.
The paper ends with Conclusion.

\section{Positivity constraints}
Let $F^I(s,t)$ denote
the $\pi\pi$ scattering amplitude with
isotopic spin $I$ in the
$s$ channel. In matrix notation ${\bf
F}(s,t)$ satisfies the
following crossing relation \cite{Ro}
$${\bf
F}(s,t)=C_{st}{\bf F}(t,s)=C_{su}{\bf F}(u,t)$$ where
the
notations are

$$ {\bf F }(s,t)=\left(\begin{array}{l}
F^0(s,t)\\
F^1(s,t)\\F^2(s,t) \end{array}\right)$$

$$C_{st}=\left( \begin{array}{lcr}1/3& 1& 5/3\\1/3&1/2&-5/6\\
1/3&-1/2&1/6\end{array} \right) \,,\,
 C_{su}=
\left(
\begin{array}{ccc}
1/3&-1&5/3\\
-1/3&1/2&5/6\\
1/3&1/2&1/6\end{array}
\right) $$
From the results of axiomatic field
theory we know that the amplitudes $F^{I}(s,t)$
satisfy fixed-t dispersion
relations with two subtractions  \cite{JM} for
$|t|<4$. We may
write them as

$${\bf F}(s,t)=C_{st}[{\bf a}(t)+(s-u){\bf
b}
(t)]+{1\over\pi}\int_4^{\infty}
{d\, x\over
x^2}{\left({s^2\over{x-s}}+{u^2\over{x-u}}C_{su}\right)\,{\bf
A}
(x,t)} \eqno(2.1)$$
where ${\bf A}(x,t)$ is the absorptive part
of ${\bf F}(s,t)$ and the
subtraction constants are of the
form
$${\bf a}(t)=\left(\begin{array}{c}a^0(t)\\ 0\\a^2(t)

\end{array}\right)\,\,\quad{\bf b}(t)=\left(\begin{array}{c} 0\\
b^1(t)\\0
 \end{array}\right)$$
due to crossing
symmetry.

In the following we shall consider that $s,t,u$
take values in
the unphysical region $|s,t,u|<4$.  We calculate
the difference
$$ {\bf F} (s,t)-{\bf F} (s_1,t) $$ and we are
looking for those
combinations of isospin amplitudes for which
this difference does
not depend on the subtraction constants.
From Eq.(2.1) we find
$${1\over {s-s_1}}( {{\bf F}(s,t)-{\bf
F}(s_1,t)}) = 2
C_{st}\,\, {\bf b}(t) + f(A)\eqno(2.2)$$ where
$f(A)$ denotes
the complicated term containing the integration
over the
absorptive parts. The first term on the right hand side
of
Eq.(2.2) is $$C_{st}\,{\bf
b}(t)=\left(\begin{array}{c}b(t)\\
{1\over2}b(t)\\-{1\over2}b(t)
\end{array}\right)$$

The last relation shows that there are
three combinations of
isospin amplitudes for which the
difference (2.2) have no
dependence on the subtraction
constants. They are $F^0+2 F^2$,
$F^1+F^2$ and $F^0-2 F^1$ and
we shall denote them as $F_i$,
$i=1,2,3$ in this order. The
first one is the well known
$\pi^0\pi^0$ elastic amplitude. One
easily obtains from Eq.(2.2)
the
relation
$$F_i(s,t)-F_i(s_1,t)={(s-s_1)(s-u_1)\over\pi}\int_4^{\infty}{(2
x+t-4)A_i(x,t)\,
dx\over (x-s)(x-s_1)(x-u)(x-u_1)}$$ $i=1,2,3$.
From this
relation we get

$${\partial F_i(s,t)\over\partial
s}={s-u\over\pi}\int_4^{\infty}{(2
x+t-4)A_i(x,t)\, dx\over
(x-s)^2(x-u)^2}$$

Because the absorptive parts $A_1$ and
$A_2$ are positive we find that
$${1\over s-u}{\partial
F_i(s,t)\over\partial s}\ge 0\,\,\,\,\,i=1,2$$
The third
combination involves the  absorptive
parts $A^0(x,t)- 2
A^1(x,t)$ whose sign is not defined and we cannot
say anything
about the sign of the  derivatives of $F_3(s,t)$.
The precedent relations 
  show us
that on the
line
 $s=u,\,\,
F_i(s,t), \,\,
i=1,2$ 
attain their minimum
values.
Indeed we 
obtain from them  
the second 
derivatives

$${\partial^2F_i(s,t)\over\partial
s^2}={2\over\pi}\int_4^{\infty}\left({1\over(x-s)^3}+{1\over(x-u)^3}\right)A_i(x,t)
\,dx\ge0,
\,\,\, 
i=1,2 \eqno(2.3)$$
\noindent
which are positive
definite, implying
that the functions
 $F_i(s,t)$ have a
minimum on the line 
$s=u$.
From the
last relation we obtain
also
$${\partial^{2n-1}F_i(s,t)\over\partial
s^{2n-1}}={(2n-1)!\over\pi}\int_4^{\infty}\left({1\over(x-s)^{2n}}-{
1\over(x-u)^{2n}}\right)
A_i(x,t)\,dx=$$
$${(2n-1)!\,(s-u)
\over\pi}\int_4^{\infty}{\left[(x-s)^n+(x-u)^n\right]}$$
$$\left[{{(x-s)^{n-1}+(x-s)^{n-2}(x-u)+\dots(x-u)^{n-1}}\over(x-s)^{2n}(x-u)^{2n}}\right]A_i(x,t)\,dx$$
In
this way we obtain the set of positivity constraints
$${1\over
s-u}{\partial^{2n-1}F_i(s,t)\over\partial s^{2n-1}}\ge
0\,\,\,
{\rm and}\,\,\, {\partial^{2n}F_i(s,t)\over\partial
s^{2n}}\ge
0\eqno(2.4)$$ $i=1,2,\,\, n=1,2,\dots$

A first remark is the
following, if the positivity constraints
have to be fulfilled
it is sufficient to test them only on the
line $s=u$, i.e.
$2s+t-4=0$, where the functions $F_i(s,t)$ attain
their minimum
values. In this way we have only one free
parameter, $0<|s|<4$,
and on this line the odd and even
derivatives give the same
information. From the point of view of
computation it is
simpler to work with  even derivatives.

Up to now we have
obtained two constraints given by Eq.(2.3). Because we
have
three  independent isospin amplitudes it follows that we
can obtain 
another one at most.

The positivity
constraints
can be imposed even on the
isospin amplitudes themselves.  This
can be easily seen from the
relation (2.1) for $F^2(s,t)$ which
after derivation gives
$${\partial^2
F^2(s,t)\over\partial
s^2}={2\over\pi}\int_4^{\infty}dx\left[\left({1\over(x-s)^3}+
{1\over
6}{1\over(x-u)^3}\right)A^2(x,t)+\right.$$
$$\left.{1\over
3}{1\over(x-u)^3}A^0(x,t)+
{1\over
2}{1\over(x-u)^3}A^1(x,t)\right]$$
 The right hand side
of the
previous relation is a positive quantity and by
iteration we
obtain that $${\partial^{2n} F^2(s,t)\over\partial
s^{2n}}\ge 0
\qquad n=1,2,\dots\eqno(2.5)$$

Unfortunately the numerical calculations show that
this relation is not 
independent of the previous two
ones.
An other
 way to obtain them is to make use of
the
Gribov-Froissart representation for the partial wave
amplitudes.
One writes dispersion relations for the isospin
amplitudes, the
subtraction constants being given by the $s$-
and $p$-wave
partial amplitudes, and one finds
$$F^I(s,t)=f_0^I(s)
+{1\over\pi}\int_0^{\infty}A^I(x,s)g(x,s,t)\,
dx\qquad
I=0,2\eqno(2.6) $$ where $$g(x,s,t)={1\over
x-t}+{1\over
x-u}+{2\over 4-s}ln(1+{s-4\over x})$$ For $I=1$ we
can write
similarly $$F^1(s,t)=3f_1^1(s)(1+{2t\over
s-4})
+{1\over\pi}\int_0^{\infty}A^1(x,s)h(x,s,t)\, dx
\eqno(2.7) $$
where $$h(x,s,t)={1\over x-t}-{1\over
x-u}+{6(2t+s-4)\over
(4-s)^2}\left[{2x+s-4\over
4-s}ln(1+{s-4\over x})+2\right]$$ 
The
absorptive parts entering
(2.6)-(2.7) are the t-channel ones and
in the following we make
use of the $t\longleftrightarrow u$ crossing
symmetry.
 From
the relation (2.6) we find analogous formulas  to (2.4),
namely
$${1\over
t-u}{\partial^{2n-1}F^I(s,t)\over\partial
t^{2n-1}}\ge 0\,\,\,
{\rm and}\,\,\,
{\partial^{2n}F^I(s,t)\over\partial t^{2n}}\ge
0\eqno(2.8)$$
$ n=1,2,\dots,\,\,I=0,2$, which  are not numerically
independent of the previous ones.
More interesting is the
relation (2.7)
which  can be written as
$${F^1(s,t)\over
t-u}
= \widehat{F^1}(s,t)= {3f_1^1(s)\over s-4}
+
{1\over\pi}\int_0^{\infty}A^1(x,s) 
\left[{1\over
(x-t)(x-u)}+\right.$$
$$\left.
{6\over
(4-s)^2}\left(
{2x+s-4\over 4-s}
ln(1+{s-4\over
x})+2\right)\right]  \,dx $$
This relation provides us with
another independent relation. Because $F^1(s,t)$ 
is
antisymmetric in $t\longleftrightarrow u$,
$\widehat{F^1}(s,t)$
is an analytic function within the Mandelstam triangle.
Deriving it with respect of $t$ we obtain

$${\partial\widehat{F^1}(s,t)\over\partial
t}={t-u\over\pi}\int_0^{\infty}{A^1(x,s)\over(x-t)^2(x-u)^2}\,
dx $$
From this relation we obtain  similar formulas to
Eq.(2.4), namely
$$
{1\over 
t-u}{\partial^{2n-1}\widehat{F^I(s,t)}\over\partial  t^{2n-1}}\ge 0\quad
{\rm and}\quad {\partial^{2n}\widehat{F^I(s,t)}\over\partial  t^{2n}}\ge 0\eqno(2.9)$$
Finally the positivity may also be
expressed as the positivity
of the partial wave amplitudes
$$f_l^I(s)={1\over
4-s}\int_4^{\infty}A^I(x,s)Q_l({2x\over
4-s}-1)\, dx\eqno(2.10)$$
for $l\ge 2$ inside the unphysical
region $0\le s\le 4$, but  the numerical calculations show that these
last constraints are weaker than those derived above.

 \section{Numerical Results}

In the previous section we derived a complete set of positivity
constraints. In an exact theory many of them are consequence of
the others as we will see later. Because ChPT does not
completely specify the amplitude and on the other hand the power
of different constraints is not the same it is useful to derive
as many
constraints as possible. Since all of them have to be
satisfied we will select the
strongest one in every case.

 To test the method  we worked first in the one-loop
approximation, ${\cal O}(p^4)$, i.e. we retained terms up to $a^2$
in Eq.(1.1). In
this order the obtained constraints do not
depend on the value
taken by $a$; in the two-loop approximation
the constraints will be linear in
$a$. We consider first the 
constraints on the $\pi^0\pi^0$ 
scattering
amplitude.

 The first two constraints (1.2) on the
$\pi^0\pi^0$, $s$-wave  $f_0(4)>f_0(0)>f_0(3.15)$
are equivalent
to $$b_3+3 b_4\ge-{9\over 1024}-{29\over
384\pi^2}\approx
-1.64\times10^{-2}\quad {\rm and}\quad
b_3+3
b_4\ge-1.47\times10^{-2}$$ respectively. The third
relation $
f_0(0)>{1\over 2}\int_2^4f_0(s)\, ds$ is equivalent
to $$b_3+3
b_4\ge -4.64\times10^{-4}$$ the strongest result
being the last
one. Already the last relation (1.2) furnishes a
better result,
the above combination of coefficients gets
positive $$b_3+3
b_4\ge 7.36\times10^{-4}$$

The derivative
of the $s$-wave has the form $$f_0'(s)={2\over
3}(b_3+3
b_4)(5s-8)+h(s)$$ where the function $h(s)$ has a
long
expression which we do not write it here.  Since $5s-8<0$,
for
$s<1.6$, the upper and lower bounds on the derivative
$f_0'(s)$
describing the position of the minimum are equivalent
to a
single lower bound on the combination $b_3+3 b_4$. The
best
result is obtained at $s=1.696587$ and is
 $$b_3+3
b_4\ge
6.67\times10^{-3}$$

The last result is the strongest
constraint upon the combination
$b_3+3 b_4$ obtained from
inequalities satisfied by the
$\pi^0\pi^0$ \, $s$-wave within
the unphysical region $0\le s\le
4$. We have given all the
above results to understand why these
constraints  were easily
satisfied by the phenomenological
models for partial-wave
amplitudes constructed in the past
years; satisfying a few of
them the others are automatically
fulfilled.

A similar analysis with analogous results was done in ref.\cite{An}
including in the game the s- ans d-waves.

Stronger constraints are
obtained from the positivity of the
second derivative of the
full amplitudes.  From Eq.(2.3) we find
for $i=1$, i.e. the
$\pi^0\pi^0$ amplitude again, a relation of
the form $$b_3+3
b_4 +h_1(s)\ge 0$$ where $h_1(s)$ is a
decreasing function for
$s<4$. Thus a problem arises, at what
point has to be
considered the above relation. We decided to
limit the range of
$s$ within the interval $|s|<4$ since we are
aware that the
amplitude (1.1) is only an approximation of the
true amplitude,
approximation truly not valid for values $s>16$
in the physical
region. $s=-4$ is equivalent to $t=12$ in the
physical region
of the $t$-channel.  One gets
 $$ b_3+3 b_4\ge {37\over
1920\pi^2}\approx 1.92\times 10^{-3}\,,
 \quad{\rm for }\quad
s=0\quad
{\rm and}$$
$$ b_3+3 b_4\ge 8.28\times
10^{-3}\,\,\quad {\rm for}\quad s=-4$$ 
which is stronger than
the previous inequality. In the following we will
list
numerical values only at $s=0$,  the numerical values at
$s=-4$
being  not very
different although they are a little
better such as the previous
 relations show.

We will work
now in the   two-loops approximation, ${\cal O}(p^6)$, and consider only
the
constraints derived in Section 2 which give the strongest
results.
 The constraints have the
form
$$\sum_{i=1}^6\,c_i^k(s,a)\,b_i+f_k(s,a)\ge 0$$
where
$k=1,2,3$  labels the independent constraints. As
we said before there are only three possible constraints
 from the positivity of the
absorptive parts,
because we have only three independent
amplitudes and the crossing symmetry
is an exact symmetry for
the ChPT amplitudes. In the previous section we
derived six
constraints but only three are numerically independent.
The
above inequalities are obtained for
every value of $s$ within the unphysical
region $0\le |s|\le
4$. Each inequality defines a half space where can live
the
parameters $b_i$. Thus the true result would be that
obtained by constructing
the intersection of these half spaces.
Unfortunately the envelope cannot
be found in an  analytic form  the
functions $f_k(s,a)$ being very complicated. 

The constraint (2.3) for $i=1$ and $s=0$ is equivalent
to
$$b_3+3 b_4-{37\over 1920\pi^2}+a\left[{7 b_1\over 320\pi^2
}-{b_2\over
60\pi^2}-{2 b_3\over 45\pi^2}+{b_4\over
180\pi^2}+16 b_6-\right.$$
$$\left.{367\over
552960\pi^2}+ {6869\over
1658880\pi^4}\right]\ge\,0\eqno(3.1)$$
 Making
$a\rightarrow 0$ one
gets the one-loop result. 

  For $i=2$ we find the
relation
$$b_4-{31\over 5760\pi^2}-a\left[{b_1\over
240\pi^2}+{43 b_2\over 2880\pi^2}+
{b_3\over 24\pi^2}+{23
b_4\over 180\pi^2}-4 b_6-\right.$$
$$\left.{67\over
276480\pi^2}-{707\over
331776\pi^4}\right]\ge 0\eqno(3.2)$$

The inequality (2.5) for $n=2$
gives
$$b_3+5\,b_4-{173\over 5760\pi^2}+a\left[{13\,b_1\over
960\pi^2}-{67\,b_2\over
1440\pi^2}-{23\,b_3\over
180\pi^2}-{b_4\over
4\pi^2}+24\,b_6-\right.$$
$$\left.{11\over
61440\pi^2}+{13939\over
1658880\pi^4}\right]\ge 0$$
 and it is easily seen that
it is
a consequence of the previous two ones being the sum of the
first and of the second one multiplied by $2$. The relations
(2.6)-(2.8) give
us, in principle, three new inequalities, but
only one will be independent of
the first two already obtained.
For $I=0$ we obtain
$$b_3+7\,b_4-{47\over
1152\pi^2}+a\left[{b_1\over
192\pi^2}-{11\,b_2\over
144\pi^2}-{19\,b_3\over
90\pi^2}-{91\,b_4\over
180\pi^2}+32\,b_6+\right.$$
$$\left. {169\over
552960\pi^2}+{7003\over
550960\pi^4}\right]\ge 0$$
which again is a linear combination
of the first two ones.

For $I=1$ the relation has the
form
$${11\over 2688\pi^2}+a\left[{b_1\over 224\pi^2}+{17
b_2\over 1344\pi^2}-
{151 b_3\over 2016\pi^2}-{653 b_4\over
2016\pi^2} +b_5+b_6+\right.$$
$$\left. {37\over 215040\pi^2}+
{4111\over
290304\pi^4}\right]\ge 0\eqno(3.3)$$

This is the third independent
relation as can  easily be seen  because the one-loop
approximation gives a positive number independent of $b_i$. More
important
is the fact that the  function $f_3(t,a)$ appearing
in the relation
(3.1) in the one loop approximation is  positive  over
(presumably) all
 the negative real axis which proves that
the
positivity is very well satisfied even by the lowest approximation
of  the ChPT amplitudes!

For
$I=2$ the inequality is
$$b_3+b_4-{49\over 5760\pi^2
}+a\left[{29\,b_1\over
960\pi^2}+{19\,b_2\over
1440\pi^2}+{7\,b_3\over
180\pi^2}+{47\,b_4\over
180\pi^2}+8\,b_6-\right.$$
$$\left. {127\over
110592\pi^2}-{67\over
552960\pi^4}\right]\ge 0\eqno(3.4)$$
We have written all the inequalities
since they were useful in checking the calculations.

A first remark is the following, the coefficient $b_5$ appears only in the inequality (3.3). Thus we can say that it is unimportant and make it vanish also in the amplitudes! It is true that
the above inequalities have been obtained by an extremal property, they are calculated on the line where the corresponding amplitudes are taking their  minimal values. This may be a suggestion that the physical partial waves satisfy also an extremal princ
iple which has to be found. This is also supported by the findings of Wanders, who could not obtain a reliable value for this parameter\cite{Wa}. What the above results suggest is that a good  determination of $b_5$ can be obtained 
 only from $I=1$ data.

We have tested how the parameters $b_i$ found in literature compare with the inequalities. Unfortunately there are only two papers that gives values for all $b_i$ \cite{Bi,A}.  The values given by Bijnens {\it et al.} are in the domain allowed by Eqs.(3.1
)-(3.4), but the values obtained by Ananthanaryan strongly violate the inequality (3.4), the previous ones being satisfied.  The values obtained by Knecht {\it et al.}\cite{Kn} for the last four parameters can be used to obtain constraints on the $b_1$ an
d $b_2$ but the  allowed domain is rather large;  the same is true for the values given by Wanders\cite{Wa}.

Using Roy equation analysis of the available $\pi\pi$ phase shift data Ananthanaryan and B\"uttiker\cite{AB} obtained values for the chiral coupling constants $\overline{l}_{1},\,\overline{l}_{2}$ but their results are not easily translated into constrain
ts on the $b_i$ coefficients
We have
tested also the inequalities (2.4), (2.8), (2.9) and (2.10) for
a few values
$ n \ge 3$ and $l \ge 2$ respectively and we found
that they are very well fulfilled.

\section{Conclusion}

We have tested
the positivity properties of the ChPT pion-pion amplitudes and
we have obtained 
a number of inequalities which express this
property. We conclude that the
pion-pion amplitudes given by
the relation (1.1) satisfy this property unexpectedly well. In the
${\cal O}(p^4)$ approximation the positivity implies
 two constraints: $b_4$ is
a positive quantity and so is the combination
 $b_3+3 b_4$. As
concerns the positivity of the $I=1$ amplitude this 
 was
tested  numerically up to $t=-10^{5}$ where the derivative is
still positive. Including ${\cal O}(p^6)$ contributions one
get constraints involving all the
 six  parameters $b_i$, but
the corrections to the ${\cal O}(p^4)$ results
   are small
which support the idea that the expansion (1.1)
 is the best
candidate for the true amplitude.
It might seem surprising but we consider that the main result
of this study is the conclusion that the most important
parameters
  to be determined are
 $b_3$ and $b_4$. This is a
consequence of the good properties near
threshold of the Weinberg
 approximation together with  the
very powerful property of positivity of the scattering
amplitudes. Let us remind that this property was essential in
deriving the
 analyticity domain of pion-pion amplitudes
\cite{M7}.
 This means that  the  amplitude (1.1) in which all
but
 $b_3,\,b_4$ are  zero will give a fair description of the
low energy phenomenology
 up to about 6-700 MeV and the
contribution of the other coefficients will be
 seen at higher
energies. In conclusion a good determination of 
 the above two
parameters will be a good starting point in the comparison of the
theory
 with experiment. Work on this line is in progress.
\vskip3mm
{\bf Acknowledgements.} This work started when the author was a visitor
at the Institute of Theoretical Physics, University of Bern in the frame of the
Swiss National Science Foundation Program. I take this
 opportunity to thank SNSF 
for support. It is a pleasure to thank  Prof. H. Leutwyler and Prof. G.
Wanders for discussions.

%\bye
%\end
\end{document}